\documentclass[%
 reprint,
superscriptaddress,
 amsmath,amssymb, aps,
]{revtex4-2}

\usepackage{graphicx}
\usepackage{dcolumn}
\usepackage{bm}

\begin{document}

\preprint{APS/123-QED}

\title{Merging and band transition of bound states in the continuum in leaky-mode photonic lattices}%

\author{Sun-Goo Lee}
\email{sungooleee@gmail.com}
\affiliation{Department of Data Information and Physics, Kongju National University, Gongju, 32588, Republic of Korea}%
\affiliation{Institute of Application and Fusion for Light, Kongju National University, Cheonan, 31080, Republic of Korea}%

\author{Seong-Han Kim}
\affiliation{Advanced Photonics Research Institute, Gwangju Institute of Science and Technology, Gwangju 61005, Republic of Korea}%

\author{Wook-Jae Lee}
\affiliation{Department of Data Information and Physics, Kongju National University, Gongju, 32588, Republic of Korea}%
\affiliation{Institute of Application and Fusion for Light, Kongju National University, Cheonan, 31080, Republic of Korea}%

\date{\today}

\begin{abstract}
Bound states in the continuum (BICs) theoretically have the ability to confine electromagnetic waves in limited regions with infinite radiative quality ($Q$) factors. However, in practical experiments, resonances can only exhibit finite $Q$ factors due to unwanted scattering losses caused by fabrication imperfections. Recently, it has been shown that ultrahigh-$Q$ guided-mode resonances (GMRs), which are robust to fabrication imperfections, can be realized by merging multiple BICs in momentum space. In this study, we analytically and numerically investigate the merging and band transition of accidental BICs in planar photonic lattices. Accidental BICs can merge at the edges of the second stop band, either with or without a symmetry-protected BIC. We show that as the thickness of the photonic lattice gradually increases, the merged state of BICs transitions from the upper to the lower band edge. Using coupled-mode analysis, we present the analytical merging thickness at which multiple accidental BICs merge at the second-order $\Gamma$ point. Our coupled-mode analysis could be beneficial for achieving ultrahigh-$Q$ GMRs in various photonic lattices composed of materials with different dielectric constants.
\end{abstract}
\maketitle



\section{Introduction}
In photonics, bound states in the continuum (BICs) refer to the special eigensolutions of electromagnetic wave equations in non-Hermitian physical systems \cite{Marinica2008,Plotnik2011,Koshelev2019}. Unusually localized BICs can exhibit theoretically infinite radiative $Q$ factors, even though they can couple with outgoing radiative waves that can carry electromagnetic energy \cite{Hsu2016}. Due to their remarkable ability to increase light-matter interactions in confined regions, extensive studies on  BICs have been conducted for basic research and practical applications in recent years \cite{YXGao2017,Minkov2018,XGao2019,SGLee2019-1,Kodigala2017,STHa2018,XYin2023}. In theoretical or numerical studies, BICs with infinite $Q$ factors can be found in various open photonic systems, including photonic crystals \cite{JWang2020,MMinkov2019}, metasurfaces \cite{Koshelev2018,Koshelev2018,Kupriianov2019}, and plasmonic structures \cite{Azzam2018,BinAlam2021,YZhou2022}. In experimental implementations, however, unwanted scattering losses attributed to fabrication imperfections, such as surface roughness, imperfect boundaries of individual scattering elements, and structural disorder, give rise to the considerable reduction of radiative $Q$ factors. In practical photonic structures, the theoretically perfect BICs appear as quasi-BICs with finite $Q$ factors due to out-of-plane coupling with continuous radiating waves.

Subwavelength photonic crystal slabs can exhibit abundant BICs if the lattices possess time-reversal symmetry, up-down mirror symmetry, and proper rotational symmetry \cite{LNi2016,SGLee2020-1}. Recently, topologically-protected BICs in photonic crystal slabs have attracted particular interest because they are stable \cite{XGao2016,Gansch2016,Yang2014,SGLee2021-1}. Furthermore, current nanofabrication technology is sufficient to implement BICs in photnoic crystal slabs \cite{Doeleman2018,BWang2020,XYin2020}. Due to their topological nature, accidental BICs can be moved along the band diagram by varying the structural parameters while maintaining the symmetry of the systems \cite{BZhen2014,MKang2022-2}. It has also been demonstrated  that multiple accidental BICs can be merged at the Brillouin zone center, which is also referred to as the lattice $\Gamma$ point in a photonic band structure, by adjusting structural parameters \cite{JJin2019,MKang2022-1}. In the merged states of multiple BICs, undesired out-of-plane scattering losses can be significantly suppressed by increasing the radiative $Q$ factors of nearby eigenstates over a finite range of wavevectors. Merging BICs is important for practical applications because it provides a powerful mechanism to achieve robust ultrahigh-$Q$ resonances, that enhance light-matter interactions.

\begin{figure*}[]
\centering\includegraphics[width=16cm]{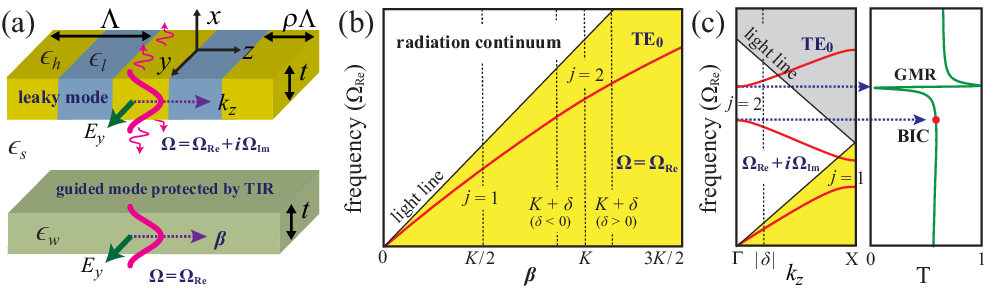}
\caption {\label{fig1} (a) Schematics of a 1D photonic lattice and homogeneous dielectric waveguide. Leaky modes in the photonic lattice lose their electromagnetic energy over time, while nonleaky guided modes in the homogeneous waveguide are protected by total internal reflection. (b) The dispersion curve for the fundamental $\mathrm{TE}_{0}$ mode of a homogeneous dielectric waveguide. (c) Conceptual illustration of photonic band structures and transmission curve through a 1D photonic lattice. Leaky modes of the photonic lattice in the lower (upper) dispersion curve at the Bloch wavevector $k_z=|\delta|$ around the second band gap ($j=2$) originate from the propagating modes of the homogeneous waveguide at the wavevector $\beta = K-|\delta|$ ($\beta = K+|\delta|$). At the second stop band, one band edge mode generate a GMR via coupling with normally incident plane wave, whereas the other edge mode becomes a symmetry-protected BIC.}
\end{figure*}

The concept of merging multiple BICs has been introduced recently and has been investigated in only a few studies so far. In this paper, we present analytical and numerical results on the merging and band transition of accidental BICs in one-dimensional (1D) and two-dimensional (2D) photonic lattice slabs. Guided modes become BICs when out-of-plane radiation completely disappears. We present an analytical expression, an overlap integral in the slab region, associated with the out-of-plane radiation of the guide modes, and show that accidental BICs can appear at generic $k$ points in reciprocal space where the value of the overlap integral approaches zero. As the thickness of photonic lattice slabs increases, accidental BICs gradually move downward along the upper dispersion curve and eventually meet at the lattice $\Gamma$ point. As the thickness is further increased, the merged state of BICs transitions from the upper to lower band edges at the lattice $\Gamma$ point. The BICs move down and away from each other with further increase in thickness. Merging thicknesses, at which multiple accidental BICs merge at the second-order $\Gamma$ point, are calculated through the coupled-mode analysis. Our coupled-mode analysis could be applied to achieve ultrahigh-$Q$ resonances in diverse 1D and 2D photonic lattices made of materials with different dielectric constants.

\begin{figure*}[]
\centering
\includegraphics[width=17.5 cm]{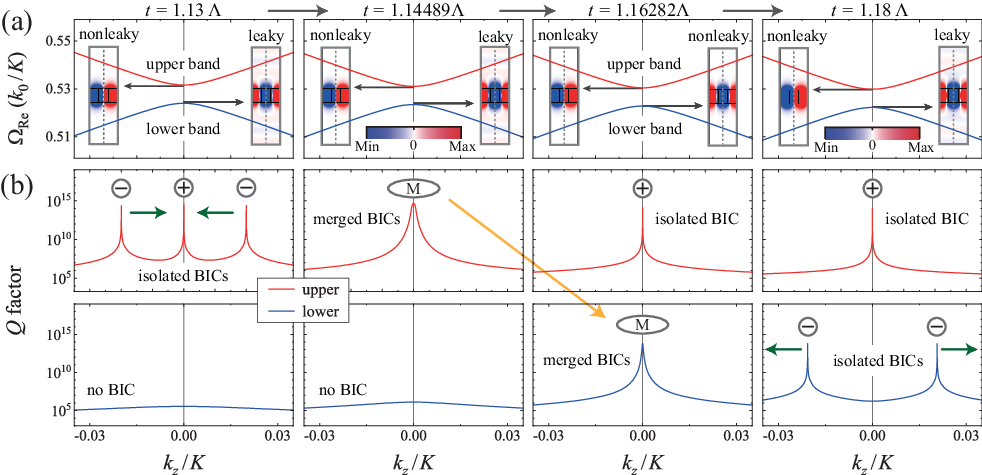}
\caption {\label{fig2} Merging and band transition of accidental BICs in a 1D leaky-mode photonic lattice. (a) FEM-simulated dispersion curves near the second stop band for four different values of $t$. Insets with blue and red colors show the spatial distribution of the electric field ($E_y$) for the band edge modes at the $y=0$ plane. The vertical dotted lines indicate the mirror planes in the computational cells. (b) Simulated radiative $Q$ factors of the upper and lower dispersion curves. The merged state of BICs transitions from the upper to lower dispersion curve as $t$ increases, while one symmetry-protected BIC with $+1$ charge always appears at the upper band edge. The structural parameters used in the FEM simulations were $\epsilon_{0}=4$, $\Delta\epsilon=0.5$, $\epsilon_{s}=1$, and $\rho = 0.35$.}
\end{figure*}

\section{Merging and band transition of BICs in 1D photonic lattices}
Figure~\ref{fig1}(a) compares a conventional homogeneous slab waveguide and a representative 1D photonic lattice for studying the merging and band transition of BICs in this study. The waveguide and photonic lattice are surrounded by a surrounding medium with a dielectric constant $\epsilon_s$. The waveguide consists of a homogeneous material with a dielectric constant $\epsilon_w$, and the 1D photonic lattice is composed of high ($\epsilon_h$) and low ($\epsilon_l$) dielectric constant materials. The period is $\Lambda$, the width of the high dielectric constant part is $\rho\Lambda$, and the thickness of the photonic lattice is $t$. In the homogeneous waveguide with $\epsilon_w > \epsilon_s$, guided modes posses purely real eigenfrequencies $\Omega = \Omega_{\mathrm{Re}}$ and propagate along the waveguide without out-of-plane radiation because they are perfectly protected by the total internal reflection (TIR). As illustrated in Fig.~\ref{fig1}(b), dispersion curves for waveguide modes should be located in yellow region below the light line. The 1D lattice can also support transverse electric ($\mathrm{TE}$) guided modes because its average dielectric constant $\epsilon_{0}=\epsilon_{l}+\rho (\epsilon_{h}-\epsilon_{l})$ is larger than $\epsilon_s$. With $\Delta\epsilon = \epsilon_{h}-\epsilon_{l} > 0$, as illustrated in Fig.~\ref{fig1}(c), all Bloch guided modes can be plotted in the irreducible Brillouin zone and photonic band gaps open at the Bragg condition $k_z = jK/2$, where $k_z$ is the Bloch wavevector, $K=2 \pi/\Lambda$ is the magnitude of the grating vector, and $j$ is an integer. Near the second stop band in the white region, guided modes posses complex eigenfrequencies $\Omega = \Omega_{\mathrm{Re}}+i~\Omega_{\mathrm{Im}}$ and exhibit interesting leaky-wave effects such as BICs and guided-mode resonances (GMRs), also known as guided resonances, via zero-order diffraction \cite{YDing2007,SGLee2019-2}. Guided modes in the yellow region are not associated with the leaky-wave effects because they do not couple with external waves in the radiation continuum, and Bloch modes in the grey region are less practical because they generate unwanted higher-order diffracted waves as well as the desired zero-order diffraction. In general, numerous leaky guided modes coexist in the photonic lattices with a slab geometry and the various modes possess their own dispersion curves, photonic band gaps, and high-$Q$ BICs. In this study, we focus our attention to the BICs in the vicinity of the second stop band ($j=2$) open by fundamental $\mathrm{TE}_{0}$ mode, as this simplest case brings out the key properties of the accidental BICs. In fact, most of important studies on the BICs are associated the leaky Bloch modes in the vicinities of the second stop bands. To elucidate the fundamental properties of BICs, we employ a semianalytical dispersion model and rigorous finite-element method (FEM) simulations.

Merging and band transition of BICs are possible because they are topologically protected \cite{BZhen2014,MKang2022-2}. Figure~\ref{fig2}(a) illustrates the evolution of the dispersion curves of guided modes in the vicinity of the second stop band under variation of the thickness $t$. No noticeable changes in the dispersion curves are observed with variations in thickness. The simulated spatial electric field ($E_y$) distributions in the insets show that regardless of $t$, upper band edge modes with asymmetric field distributions become nonleaky symmetry-protected BICs. Conversely, lower band edge modes with symmetric field distributions can be either leaky or nonleaky depending on the thickness values. The merging and band transition of accidental BICs can be clearly seen in Fig.~\ref{fig2}(b), which depicts $Q$ factors as a function of $k_z$. When $t=1.13~\Lambda$, two accidental BICs with $-1$ charge and one symmetry-protected BIC with $+1$ charge appear, resulting in three isolated peaks in the $Q$ factor curve of the upper band. Due to the $180^\circ$ rotational symmetry of the 1D lattice, two accidental BICs appear in pairs at wavevectors $k_z=\pm 0.01995~K$ simultaneously. As the value of $t$ increases from $1.13~\Lambda$, two accidental BICs gradually move downward along the upper dispersion curve and approach to $\Gamma$ point where a symmetry-protected BIC is located. When $t=1.14489~\Lambda$, two accidental BICs and one symmetry-protected BIC merge at the lattice $\Gamma$ point, resulting in a single enhanced peak in the $Q$ factor curve. Compared to the isolated symmetry-protected or accidental BICs, the merged state at $t=1.14489~\Lambda$ exhibits significantly improved radiative $Q$ factors over a wide range of wavevectors. As the value of $t$ further increases to $1.16282~\Lambda$, simultaneous peaks in the $Q$ factor curves are observed at both the upper and lower band edges. The peak in the upper band is attributed to the symmetry-protected BIC, evident from the asymmetric field distributions shown in the inset of Fig.~\ref{fig2}(a). We propose that the interband transition of the merged state of BICs, induced by the variation of $t$, leads to the enhanced peak at the lower band edge. Upon further increasing $t$ beyond $1.16282~\Lambda$, the merged state at the $\Gamma$ point splits into two isolated peaks with $-1$ charge.

\section{Coupled-mode anaylsis}
In the planar waveguide structures shown in Fig.~\ref{fig1}(a), the dispersion relations of guided modes, including eigenfrequencies and radiative $Q$ factors, can be obtained by solving the 1D wave equation given by \cite{Yariv1984}
\begin{equation}\label{wave-equation}
\left (\frac{\partial^{2}}{\partial x^{2}} + \frac{\partial^{2}}{\partial z^{2}} \right ) E_{y}(x,z) + \epsilon (x,z) k_{0}^2 E_{y}(x,z)= 0,
\end{equation} 	 	
where $k_{0}$ denotes the wave number in free space. For the homogeneous waveguide plotted in Fig.~\ref{fig1}(b), Eq.~(\ref{wave-equation}) can be solved analytically and spatial electric field distribution of the $\mathrm{TE}_{0}$ mode satisfies $E_x=0$, $E_z=0$, and
\begin{equation}\label{waveguide-mode}
E_{y}(x,z) = \varphi(x) e^{i\beta z},
\end{equation} 	 	
where $\varphi(x)$ characterizes the transverse profile of the $\mathrm{TE}_{0}$ mode and $\beta$ denotes the propagation constant along the $z$ direction \cite{Agrawal2004}. The $\mathrm{TE}_{0}$ mode can propagate without radiative loss along the waveguide due to TIR.

For the photonic lattices in Fig.~\ref{fig1}(a), Eq.~(\ref{wave-equation}) can be solved numerically by expanding the electric field $E_{y}(x,z)$ as a Bloch form and the periodic dielectric function $\epsilon (x,z)$ in a Fourier series \cite{Inoue2004}. Using the current state of computational software and hardware, it is possible to obtain exact solutions of Eq.~(\ref{wave-equation}) for certain system parameters. However, relying solely on direct numerical simulations may not be sufficient to fully comprehend the fundamental properties of BICs, even though numerical calculations can provide accurate dispersion relations of leaky guided modes. To gain a deeper understanding of the merging and band transition of BICs, we employ the semianalytical dispersion model introduced by Kazarinov and Henry (KH) \cite{Kazarinov1985}. In this model, the spatial dielectric function and electric field are approximated as:
\begin{align}
\epsilon(x,z) &= \gamma_{0} + \sum_{n=1,2} \left(\gamma_{n}e^{inKz} + \gamma_{-n}e^{-inKz} \right), \label{KH-epsilon} \\
E_{y}(x,z) &= \left(Ae^{iKz} + Be^{-iKz}\right)\varphi(x) + \Delta E(x,z). \label{KH-efield}
\end{align}
In Eq.~(\ref{KH-epsilon}), the Fourier coefficient $\gamma_n$ has a zero value outside the periodic layer, and $\gamma_0=\epsilon_{0}$. In Eq.~(\ref{KH-efield}), $A(z)\sim\exp(ik_z z)$ and $B(z)\sim\exp(-ik_z z)$ are the slowly varying envelopes of the two counter-propagating waves, and $\Delta E(x,z)$ represents the diffracted wave radiating away from the periodic structure. For symmetric lattices with $\epsilon(x,z) = \epsilon^{\ast}(x,-z)$ and $\gamma_{-n} = \gamma_{n}$, near the second stop band, the eigenfrequencies of guided modes can be written as
\begin{equation}\label{dispersion}
\Omega(k_{z})=\Omega_{0} - \frac{ih_{1} \pm \sqrt{k_{z}^2+(h_{2}+ih_{1})^2}}{Kh_{0}},
\end{equation}
where $\Omega_{0}$ is the Bragg frequency under vanishing index modulation, and the coupling coefficients $h_{n=0,1,2}$ are given by
\begin{eqnarray}
h_{0}&=& \frac{k_0}{K} \int_{-\infty}^{\infty} {\gamma_{0}}(x) \varphi(x)\varphi^*(x) dx, \label{h0}\\
h_{1}&=&i \frac{ k_0^4 \gamma_1^2}{2K} \int_{-t}^{0} \int_{-t}^{0} G(x,x^{\prime})\varphi(x')\varphi^*(x) dx' dx, \label{h1}\\
h_{2}&=& \frac{k_0^2 \gamma_2}{2K} \int_{-t}^{0} \varphi(x)\varphi^*(x) dx, \label{h2}
\end{eqnarray}
where $G(x,x')$ denotes the Green's function for diffracted fields \cite{SGLee2019-1,YDing2007}. Equation~(\ref{dispersion}) indicates that the leaky stop band with two band edges $\Omega^b=\Omega_{0}+h_{2}/(Kh_{0})$ and $\Omega^a=\Omega_{0}-(h_{2}+i2h_{1})/(Kh_{0})$ opens at $k_{z} = 0$. For symmetric lattices, the coupling coefficient $h_1$ is generally a complex value, while $h_0$ and $h_2$ are real values, irrespective of the lattice parameters. With the time dependence of $\exp(-i\Omega t)$, Bloch modes near the second stop band tend to lose their electromagnetic energy over time because they have complex frequencies. However, one of the band edge modes with a purely real eigenfrequency $\Omega^b$ can become a nonleaky symmetry-protected BIC at the lattice's $\Gamma$ point.

By investigating the dispersion relations in Eq.~(\ref{dispersion}) with the associated coupling coefficients $h_0$, $h_1$, and $h_2$, one can notice that $\mathrm{Im}[\Omega]$ becomes zero when $\mathrm{Re}[h_1]$ approaches zero. With $\mathrm{Re}[h_1]=0$, dispersion relations in Eq.~(\ref{dispersion}) can be rewritten as
\begin{equation}\label{dispersion-BIC}
\Omega(k_{z})=\Omega_{0} + \frac{\mathrm{Im}[h_{1}] \pm \sqrt{k_{z}^2+(h_{2} - \mathrm{Im}[h_{1}])^2}}{Kh_{0}}.
\end{equation}
Equation (\ref{dispersion-BIC}) shows that accidental BICs with purely real eigenfrequencies can be observed at any generic $k$ point, including the lattice $\Gamma$ point. Due to their topological nature, these accidental BICs do not destroy but rather move along dispersion curves as the lattice parameters change continuously. In 1D photonic lattices possessing $180^\circ$ rotational symmetry, two accidental BICs should appear in pairs at $k_z = \pm |k_{\text{BIC}}|$ simultaneously. It is important to note that the formation of accidental BICs is independent of the existence of symmetry-protected BICs. Thus, at the asymmetric edge, two accidental BICs and one symmetry-protected BIC can merge as illustrated in Fig.~\ref{fig2} when $\mathrm{Re}[h_1]$ approaches 0 due to a variation in lattice parameters. On the other hand, at the symmetric edge, two accidental BICs can merge with the appropriate lattice parameters. In previous studies \cite{SGLee2019-1,SGLee2019-2}, it was shown that the position of a symmetry-protected BIC changes from the upper to lower band edge through the band flip, also known as the topological phase transition. This transition can be achieved by varying the value of the parameter $\rho$. Before the band flip, as shown in Fig.~\ref{fig2} with $\rho=0.35$, accidental BICs can merge with a symmetry-protected BIC at the upper band edge. Conversely, after the band flip with appropriate value of $\rho$, the merged state with a symmetry-protected BIC appears at the lower band edge.

\begin{figure}[]
\centering \includegraphics[width=8.0 cm]{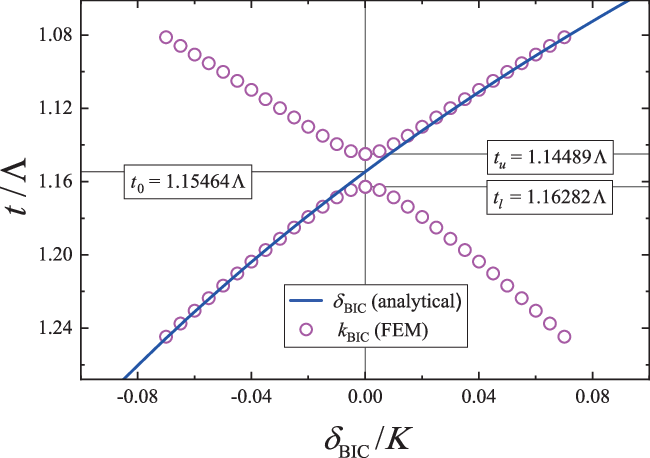}
\caption {\label{fig3} The locations of the accidental BICs, $\delta_{\text{BIC}}$, represented by a blue solid line as a function of $t$, are obtained from the 1D coupled-mode analysis. The overlap integral $\Sigma$ in Eq.~(\ref{overlap}) becomes zero at $\delta_{\text{BIC}}$. FEM-simulated BIC points, $k_{\text{BIC}}$, denoted by magenta circles, are also plotted for comparison. In the calculation of $\delta_{\text{BIC}}$, $\epsilon_{0}$ and $\epsilon_{s}$ are fixed to 4 and 1, respectively, so that the analytically obtained $\delta_{\text{BIC}}$ explains the evolution of the accidental BICs simulated by FEM in Fig.~\ref{fig1}. The parameter $t$ was used as the vertical axis, such that BIC points move downward as $t$ increases. Since the eigenfrequencies of guided modes get lower as $t$ increases, the FEM-simulated magenta circles mimic the dispersion curves represented in Fig.~\ref{fig2}(a).}
\end{figure}

We now present an analytical expression that is associated with the formation and band transition of accidental BICs. By examining the value of $h_1$ with the appropriate Green's function \cite{SGLee2019-1}, we can demonstrate that  $\mathrm{Re}(h_{1})$ approaches zero when the overlap integral given by
\begin{equation}\label{overlap}
\Sigma =\int_{-t}^{0} e^{i\mu x} \varphi(x) dx,
\end{equation}
approaches zero; see Supplemental Material \cite{Supplemental} for details. In Eq.~(\ref{overlap}), $\mu = \sqrt{\epsilon_0 k_0^2 - \delta^2}$, and the transverse profile of the $\mathrm{TE}_{0}$ mode is given by $\varphi(x)= c e^{+i \alpha x} + c^* e^{-i \alpha x}$, where $\alpha =\sqrt{\epsilon_0 k_0^2 - \beta^2}$, and the complex coefficient $c$ is set to satisfy $\int_{-\infty}^{\infty} |\varphi(x)|^2dx = 1$. As shown in Figs.~\ref{fig1}(b) and \ref{fig1}(c), waveguide modes with a positive (negative) value of $\delta$ at $\beta = K + \delta$ correspond to the leaky modes in the upper (lower) dispersion curve of photonic lattices at the Bloch wavevector $k_z=|\delta|$. Merging and band transition of accidental BICs illustrated in Fig.~\ref{fig2} can be explained by investigating analytical BIC points $\delta_{\text{BIC}}$, where the overlap integral $\Sigma$ becomes zero, as a function of the thickness $t$. Figure~\ref{fig3} illustrates the calculated $\delta_{\text{BIC}}$ with $\Sigma=0$. In the calculations, the value of $t$ varies from $\Lambda$ to $1.3~\Lambda$, and $\beta$ changes in the small discrete step of $10^{-5}~K$. $\delta_{\text{BIC}}(t)$, represented by the blue solid line, is defined as the first value at which the value of $|\Sigma|$ becomes smaller than $10^{-5}$ for a given $t$. Figure~\ref{fig3} indicates that as the thickness gradually increases from $\Lambda$, $\delta_{\text{BIC}}$ monotonically moves downward and reaches $\delta = 0$ ($\beta = K$), which corresponds to the second-order $\Gamma$ point, when the thickness is $t_0= 1.15464~\Lambda$. As the thickness increases further, $\delta_{\text{BIC}}$ continues to move downward and gets away from the $\Gamma$ point.

The analytical BICs in Fig. \ref{fig3} and the dispersion curve in Fig. \ref{fig1}(b) indicate that the eigenfrequencies of accidental BICs decrease continuously with increasing thickness. However, the effects of dielectric constant modulation, such as band folding and photonic band gap, cannot be observed in the analytical BICs represented by the blue line in Fig.\ref{fig3}, as the overlap integral $\Sigma$ in Eq.(\ref{overlap}) is determined by the transverse mode profile $\varphi(x)$ and the Green's function for the unmodulated homogeneous waveguide slab of thickness $t$. To verify whether accidental BICs indeed occur as $\Sigma$ approaches zero, we conducted additional investigations of the BIC points $k_{\text{BIC}}$ using FEM simulations. The results are represented as magenta circles in Fig.~\ref{fig3}. Due to the mirror symmetry of the photonic lattice, FEM-simulated accidental BICs with infinite $Q$ factors appear in pairs at $\pm |k_{\text{BIC}}|$. When the thickness ($t < t_0$) is far from $t_0 = 1.15464~\Lambda$, the values of $\delta_{\text{BIC}}$ and $k_{\text{BIC}}$ agree well. However, as $t$ approaches $t_0$, the FEM-simulated values of $k_{\text{BIC}}$ slightly deviate from the analytic values of $\delta_{\text{BIC}}$. The deviation $|k_{\text{BIC}} - \delta_{\text{BIC}}|$ increases until $t$ approaches $t_u = 1.14489~\Lambda$, where $k_{\text{BIC}}$ reaches the $\Gamma$ point. As $t$ further increases beyond $t_u$, the BIC point is not found for a while, but when $t_l = 1.16282~\Lambda$, the BIC point appears again at the $\Gamma$ point. The deviation $|k_{\text{BIC}} - \delta_{\text{BIC}}|$ decreases with additional increases in thickness ($t > t_l$), and $k_{\text{BIC}}$ agrees well with $\delta_{\text{BIC}}$ if $t$ is far away from $t_l$. In the vicinity of the second-order $\Gamma$ point, photonic band gaps open by the coupling between two counterpropagating waves \cite{SGLee2019-1}. Since the coupling strength is maximum at $k_z=0$ and decreases rapidly as the Bloch wavevector moves away from the $\Gamma$ point, the deviation $|k_{\text{BIC}} - \delta_{\text{BIC}}|$ could be noticeable only around the $\Gamma$ point. In the FEM simulations, there exist a thickness range, $t_u < t< t_l$, where accidental BIC can not be found due to the presence of a photonic band gap. At $t=t_u$ and $t=t_l$, two accidental BICs meet at the upper and lower band edges, respectively, resulting in the formation of merged BIC states. A symmetry-protected BIC can be included in the merged state at the upper or lower band depending on the value of lattice parameter $\rho$, not $t$.

\begin{figure*}[]
\centering
\includegraphics[width=14.5 cm]{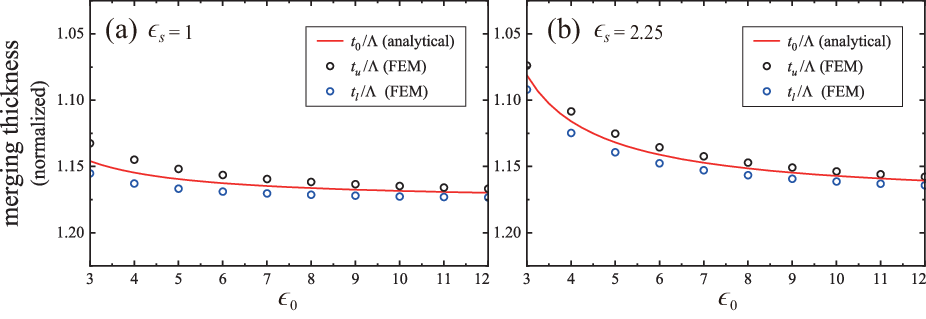}
\caption {\label{fig4} Analytical and FEM-simulated normalized merging thicknesses as a function of $\epsilon_0$ for (a) $\epsilon_s=1$ and (b) $\epsilon_s=2.25$. Analytical results $t_0/\Lambda$ are beneficial to find the merging thicknesses $t_u/\Lambda$ and $t_l/\Lambda$. In the FEM simulations, structural parameters $\rho=0.35$ and $\Delta\epsilon = 0.5$ are kept to constant. }
\end{figure*}

\section{Ultrahigh-$Q$ resonances robust to fabrication imperfections}
Our coupled-mode analysis reveals that the value of $\Sigma$ can be tuned to zero at the second-order $\Gamma$ point of photonic lattices by adjusting the lattice parameters. This tuning can cause accidental BICs to merge, resulting in interesting ultrahigh-$Q$ resonances. The overlap integral $\Sigma$ in Eq.~(\ref{overlap}) mathematically depends on the lattice parameters $t$, $\epsilon_0$, and $\epsilon_s$. However, thickness variation is essential to achieve merged states of BICs. While positions of accidental BICs can be substantially moved along dispersion curves with thickness variation, BICs only move slightly with variations of $\epsilon_0$ and $\epsilon_s$. In conventional experimental configurations at wavelengths around $\lambda = 1550$ nm, the value of $\epsilon_s$ typically ranges from 1 to 2.25, and $\epsilon_0$ varies from 3 to 12. For photonic lattices with these typical values of $\epsilon_0$ and $\epsilon_s$, it is meaningful to investigate the normalized merging thickness $t_0/\Lambda$, where $\Sigma = 0$. Red lines in Figs.~\ref{fig4}(a) and \ref{fig4}(b) display the calculated $t_0/\Lambda$ as a function of $\epsilon_0$ for $\epsilon_s=1$ and $\epsilon_s=2.25$, respectively. FEM-simulated merging thicknesses $t_u/\Lambda$ and $t_l/\Lambda$, where accidental BICs merge at the upper and lower band edges, respectively, are also displayed for comparison. Figure~\ref{fig4} demonstrates that $t_0/\Lambda$ obtained from Eq.~(\ref{overlap}) could provide useful reference for achieving merged states of BICs for ultrahigh-$Q$ resonances. In Fig.~\ref{fig4}, the values of $t_0/\Lambda$ change from $1.14588$ to $1.16981$ when $\epsilon_s = 1$, and from $1.08116$ to $1.16066$ when $\epsilon_s = 2.25$. As the value of $\epsilon_0$, representing the averaged dielectric constant in the waveguide region, decreases, the calculated and simulated merging thicknesses decrease. Figure~\ref{fig4} also shows that a higher dielectric constant $\epsilon_s$ in the cladding region results in a decrease in merging thicknesses. Changes in $\epsilon_0$ and $\epsilon_s$ have a limited impact on the analytical and FEM-simulated merging thicknesses. The dependency of merging thicknesses on the dielectric constants $\epsilon_0$ and $\epsilon_s$ may be attributed to the strong confinement of guided modes in the waveguide and the scaling property of Maxwell's equations \cite{Joannopoulos1995}.

\section{Merging and band transition of BICs in 2D photonic lattices}
\begin{figure*}[]
\centering \includegraphics[width=17.5 cm]{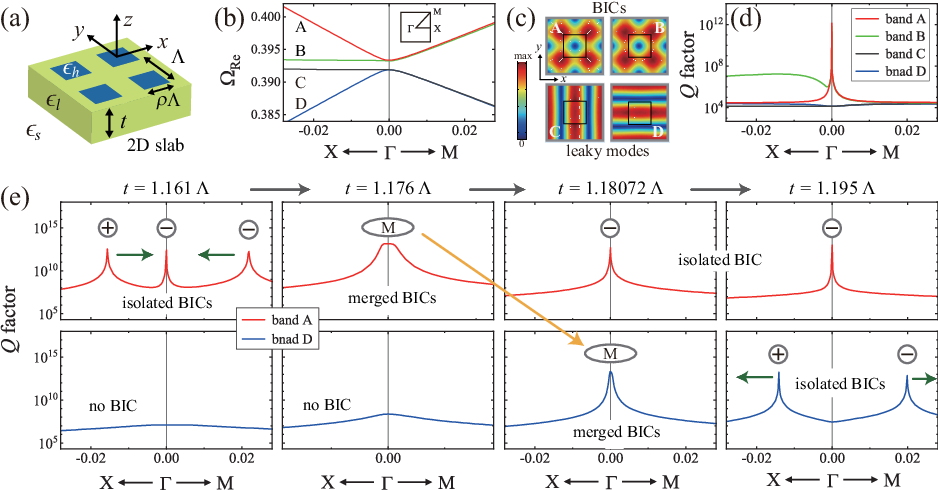}
\caption {\label{fig5} Merging and band transition of accidental BICs in a 2D photonic crystal slab. (a) Schematic of  a 2D photonic crystal slab. (b) FEM-simulated dispersion curves near the second-order $\Gamma$ point. (c) Simulated spatial electric field ($|\mathbf{E}|$) distributions of four band edge modes at $z=0$ plane. White arrows represent in-plane electric field vectors (d),(e) Simulated radiative $Q$ factors. The structural parameters used in the simulations were $\epsilon_{0}=9$, $\epsilon_{s}=1$, $\Delta\epsilon=1$, and $\rho = 0.4$. (b)--(d) were simulated with $t=0.5~\Lambda$.}
\end{figure*}

We also investigate the merging and band transition of BICs in 2D photonic crystals with a thin-film geometry. As illustrated in Fig.~\ref{fig5}(a), we use a simple 2D periodic slab composed of  square arrays of square-shaped high dielectric constant ($\epsilon_h$) materials in the background medium with a low dielectric constant ($\epsilon_l$). Here, the averaged dielectric constant is given by $\epsilon_0=\epsilon_l + \rho^2 \Delta \epsilon$. FEM-simulated dispersion curves plotted in Fig.~\ref{fig5}(b) shows that there are four bands, A, B, C, and D, in the vicinity of the second-order $\Gamma$ point. Additionally, spatial electric field ($|\mathbf{E}|$) distributions of four band edge modes and radiative $Q$ factors displayed in Figs.~\ref{fig5}(c) and \ref{fig5}(d), respectively, show that the band edge modes in A and B are nonleaky symmetry-protected BICs. Conversely, band edge mode in C and D are degenerate and generally radiative out of the photonic crystal slab. Merging and band transition of accidental BICs can be clearly seen in Fig.~\ref{fig5}(e), which depicts the evolution of radiative $Q$ factors as the slab thickness varies. The radiative $Q$ factors are simulated along the $\Gamma \mathrm{X}$ and $\Gamma \mathrm{M}$ directions in the Brillouin zone. In this study, we investigate the merging and band transition of BICs in the highest band A and the lowest band D, for convenience. When $t=1.161~\Lambda$, two accidental BICs with $\pm1$ charge and one symmetry-protected BIC with $-1$ charge appear, resulting in three distinct peaks in the $Q$ factor curve in band A. Accidental BICs with $+1$ and $-1$ charge approach the symmetry-protected BICs along the $\Gamma \mathrm{X}$ and $\Gamma \mathrm{M}$ directions, respectively. Due to the $C_4$ rotational symmetry of the 2D lattice, four sets of accidental BICs with $+1$ charge and four sets with $-1$ charge appear simultaneously. As the value of $t$ increases from $1.161~\Lambda$, eight accidental BICs gradually move downward along the band A and approach to $\Gamma$ point where a symmetry-protected BIC is located. When $t=1.176~\Lambda$, eight accidental BICs and one symmetry-protected BIC merge at the lattice $\Gamma$ point, resulting in a noticeably enhanced peak in the $Q$ factor curve. When $t=1.18072~\Lambda$, two peaks are observed in the $Q$ factor curves at the edges of band A and band D. The peak in band A is attributed to the symmetry-protected BIC, while the enhanced peak in band D arises from the merged state of accidental BICs. As $t$ is increased beyond $1.18072~\Lambda$, the merged state at the $\Gamma$ point splits into isolated peaks with $+1$ and $-1$ charges, and these peaks move downward along band D.

In the FEM simulations with 2D photonic crystal slabs, the merging of BICs was achieved when $t=1.176~\Lambda$ and $t=1.18072~\Lambda$, which are close to the analytical merging thickness of $t_0/\Lambda = 1.167$ from Fig.~\ref{fig4}(a). Furthermore, Jin \emph{et al.} have recently demonstrated experimentally that on-chip photonic resonances can be made robust against fabrication imperfections by combining nine BICs in 2D photonic crystal slabs \cite{JJin2019}. The merged state of BICs was achieved at a normalized thickness of $h/a=1.13$, which is also close to $t_0/\Lambda =1.167$. Hence, it is reasonable to conclude that the analytical merging thickness obtained from Eq.~(\ref{overlap}) could also be beneficial for achieving merged states of BICs in practical 2D photonic crystal slabs. Compared to the merged states in a 1D photonic lattice plotted in Fig.~\ref{fig2}(b), the merged states in a 2D lattice exhibit significantly improved radiative $Q$ factors across a wide range of wavevectors. This improvement is due to the convergence of a larger number of BICs at the center of the Brillouin zone in 2D configurations. Jin \emph{et al.} showed that there was an improvement in the scaling property from $Q\propto1/k_z^{-2}$ to $Q\propto1/k_z^{-6}$ by merging eight accidental BICs and one symmetry-protected BIC in 2D photonic crystal slabs. 

\section{Conclusion}
In conclusion, we analytically and numerically investigated the formation, merging, and band transition of accidental BICs in 1D and 2D photonic crystal slabs. Using a simple coupled-mode analysis, we derived an analytical expression for the overlap integral associated with the out-of-plane radiation of guided modes. Accidental BICs can emerge at generic $k$ points, including the lattice $\Gamma$ point, where the value of the overlap integral approaches zero. Due to the $180^\circ$ ($90^\circ$) rotational symmetry of 1D (2D) lattices, two (eight) accidental BICs appear in the momentum space simultaneously. As the thickness of the slab increases, the accidental BICs move downward along the dispersion curve, resulting in the merged state at the edges of the second stop band. As the thickness increases further, the merged state of BICs transitions from the upper to the lower band edge. With a further increase in thickness, the merged BICs at the lower band edge split into multiple isolated ones and move away from each other. We presented the theoretical merging thickness $t_0/\Lambda$, which is the ratio between the thickness and period of photonic crystal slabs, where multiple BICs merge at the second-order $\Gamma$ point. The merging of multiple BICs are important for practical applications because it provides a powerful mechanism to achieve robust ultrahigh-$Q$ resonances capable of enhancing light-matter interactions. Our coupled-mode analysis and FEM-simulated results are helpful for achieving topologically enabled ultrahigh-$Q$ resonances that are robust to fabrication imperfections in diverse 1D and 2D photonic lattices made of various dielectric materials.

\bigskip
\noindent\textbf{Supporting Information}\\
Supporting Information is available from the Wiley Online Library or from the author.

\bigskip
\noindent\textbf{Acknowledgements}\\
This research was supported by the grant from the National Research Foundation of Korea, funded by the Ministry of Science and ICT (No. 2022R1A2C1011091).

\bigskip
\noindent\textbf{Conflict of Interest}\\
The authors declare no conflicts of interest.

\bigskip
\noindent\textbf{Data Availability Statement}\\
Data underlying the results in this paper may be obtained from the authors upon reasonable request.

\bigskip
\noindent\textbf{Keywords}\\
bound states in the continuum, guided-mode resonances, topologically enabled ultrahigh-$Q$


\end{document}